\documentclass[aps,prb,twocolumn,superscriptaddress,longbibliography]{revtex4-2}
\usepackage{amsmath}
\usepackage{chemformula}
\usepackage{physics}
\usepackage{amsfonts}
\usepackage{bm}
\usepackage{graphicx}
\usepackage{hyperref}
\usepackage{pifont}
\usepackage{color}
\usepackage{tabularx}

\newcommand{\unit}[1]{\,\mathrm{#1}} 
\newcommand{\equa}[1]{Eq.~\eqref{#1}} 
\newcommand{\fig}[1]{Fig.~\ref{#1}}

\usepackage[T1]{fontenc}
\usepackage{times}{\Large}

\begin{document}

\author{Fengyuan Xuan}
\email[Contact author: ]{xuanfy@szlab.ac.cn}
\affiliation{Suzhou Laboratory, Suzhou, 215123, China}

\author{Jiexi Song}
\affiliation{Suzhou Laboratory, Suzhou, 215123, China}

\author{Zhiyuan Sun}
\email[Contact author: ]{zysun@tsinghua.edu.cn}
\affiliation{State Key Laboratory of Low-Dimensional Quantum Physics and Department of Physics, Tsinghua University, Beijing 100084, P. R. China}
\affiliation{Frontier Science Center for Quantum Information, Beijing 100084, P. R. China}

\title{\textit{Ab initio} Approach to Collective Excitations in Excitonic Insulators}

\date{\today}

\begin{abstract}
An \textit{ab initio} approach is presented for studying the collective excitations in excitonic insulators, charge/spin density waves and superconductors.
We derive the Bethe-Salpeter-Equation for the particle-hole excitations in the quasiparticle representation, from which the collective excited states are solved and the corresponding order parameter fluctuations are computed. 
This method is demonstrated numerically for the excitonic insulating phases of the biased WSe$_2$-MoSe$_2$ bilayer.
It reveals the gapless phase-mode, the subgap Bardasis-Schrieffer modes and the above-gap scattering states.
Our work paves the way for quantitative predictions of excited state phenomena from first-principles calculations in electronic systems with spontaneous symmetry breaking.
\end{abstract}

\maketitle

\textit{Ab initio} approaches based on density functional theory (DFT) combined with many-body perturbation theory (MBPT)~\cite{RMP2002} are widely used to study many-electron systems.
For systems with spontaneous symmetry breaking such as superconductors and excitonic insulators,
first-principles methods for their ground states have been established
~\cite{SC1988,SC1999,SC2005,EI2017,EI2019,WTe2-1,EI2023}.
However, their collective excitations~\cite{Phase,BaSh} are still inaccessible within the \textit{ab initio} framework.

The excitonic insulator (EI),  proposed six decades ago~\cite{Kel1965,Kohn}, has seen a revival in recent years~\cite{Kaneko2025}.
It refers to a long range ordered state formed by bounded electron-hole ($e$-$h$) pairs which spontaneously breaks either the $U(1)$ symmetry of exciton number conservation or discrete lattice symmetries~\cite{Portengen.1996,Zenker.2014,Mazza.2020,Golez.2020,Sun.2021,SZY2021,Kaneko2025}.
Recent experiments vibrantly searched for EI states in bulk materials such as Ta$_2$NiSe$_5$ \cite{TNS0,TNS1,TNS2,TNS3,TNS4,TNS5,TNS6,TNS7,TNS8}, WTe$_2$ \cite{WTe2,WTe2-3} and Ta$_2$Pd$_3$Te$_5$ \cite{Hasan2025}, and established stable exciton fluids in $e$-$h$ bilayer systems~\cite{KFM2019,KFM2021,Wang2022,Du2023,KFM2025,SJ2025}.
The collective excitations of the condensate correspond to fluctuations of the order parameter~\cite{Wu.2015,SZY2020, Murakami.2020,Xue.2020, Golez.2020,SZY2021,DX}, which govern optical responses and may serve as their experimental signatures.
Analytical approaches to these modes rely on model Hamiltonians~\cite{SZY2020,Murakami.2020,SZY2021,SZY2024,Mac2015,Mac2020,Golez2025,DX}, whereas realistic material calculations call for a computational formalism natively compatible with first-principles packages.

In this work, we present an \textit{ab initio} approach for quantitative computation of collective excitations using the Bethe-Salpeter-Equation (BSE) for typical systems with spontaneously broken symmetry.
While the presented formula apply equally well to generic systems such as superconductors and charge/spin density waves, we focus our discussion on EIs.
On top of the mean field Hamiltonian, we construct from the fluctuation terms the BSE eigenvector/eigenvalue equation for particle-hole pair excitations.
The eigenvectors and eigenvalues are those for collective excited states, from which the excitonic order parameter fluctuations are also computed.
We demonstrate our method in the biased WSe$_2$-MoSe$_2$ system and reveal the zero-energy phase-mode and a series of subgap Bardasis-Schrieffer (BaSh) modes~\cite{BaSh, SZY2020, DX}.

We start from the Hamiltonian 
that includes multiple conduction and valence bands \cite{Kohn,Kel1965,Kel1968,arXivLouie}:
\begin{eqnarray}\label{eqn:H}
\begin{split}
     H =& \sum_{ck} \xi_{ck} c^{\dagger}_{ck} c_{ck} + \sum_{vk} \xi_{vk} c^{\dagger}_{vk} c_{vk} \\
&- \sum_{vck,v'c'k'} V_{vck,v'c'k'} c^{\dagger}_{c'k'} c_{v'k'} c^{\dagger}_{vk} c_{ck}
\end{split}
\end{eqnarray}
where $\xi_{nk}$ is the single particle energy, $c$ ($v$) is an integer that runs over all conduction (valence) bands, and $V_{vck,v'c'k'}$ consists of the direct and exchange $e$-$h$ interactions.
These parameters are obtained from standard first-principles DFT and MBPT calculations \cite{QE,QE2,BGW}. 
For notational simplicity, the other two-body interaction terms such as electron-electron ($e$-$e$) and hole-hole ($h$-$h$) interactions are neglected and will be discussed later.
The $e$-$h$ pairing terms with non-zero center-of-mass momenta are not shown explicitly because we focus on zero momentum condensate and excitations in this paper.
As shown by \equa{eqn:H}, we limit our calculation to the case of conserved exciton number so that the possible EI state breaks the continuous $U(1)$ symmetry.
The effects of intrinsic interband tunneling~\cite{Portengen.1996,Sun.2021,SZY2021,SZY2024} and electron-phonon coupling~\cite{Zenker.2014,Mazza.2020,Golez.2020, Kaneko2025} will be addressed in future works.

We separate the Hamiltonian into mean-field (MF) and fluctuation parts: $H=H_{\text{MF}} +H_{\text{fluc}}$ \cite{SI,Coleman,Altland}.
The MF Hamiltonian reads
\begin{eqnarray}\label{eqn:HMF}
\begin{split}
   H_{\text{MF}} = H_0 + \sum_{ck} E_{ck} \gamma^{\dagger}_{ck} \gamma_{ck} + \sum_{vk} E_{vk} \gamma^{\dagger}_{vk} \gamma_{vk},
\end{split}
\end{eqnarray}
which is obtained from diagonalizing
\begin{eqnarray}
\begin{split}
h_k=
     \begin{bmatrix}
     ... & ... & ... & ... & ... & ... \\ 
     ... &\xi_{c_2k} & 0 & -\Delta_{v_1c_2k} & -\Delta_{v_2c_2k} & ...\\ 
     ... &0 & \xi_{c_1k} & -\Delta_{v_1c_1k} & -\Delta_{v_2c_1k}& ... \\
     ...& -\Delta^*_{v_1c_2k} & -\Delta^*_{v_1c_1k} & \xi_{v_1k} & 0 &... \\ 
     ...& -\Delta^*_{v_2c_2k} & -\Delta^*_{v_2c_1k} & 0 &  \xi_{v_2k}&... \\
     ...& ... &... & ... & ... &  ... 
     \end{bmatrix}
\end{split}
\label{SCF}
\end{eqnarray}
by the unitary matrix $U$, $U^{\dagger}h_kU=\mathrm{diag}(E_{nk})$.
Here $\Delta_{vck}$ is the band and momentum dependent excitonic order parameter.
The annihilation operators $\gamma_{ck}$ ($\gamma_{vk}$) are for the quasi-particles in the conduction (valence) bands with quasiparticle energies $E_{ck}$ ($E_{vk}$) renormalized by excitonic order. 
They are related to the bare electronic annihilation operators as $c_{nk} = \sum_m U^{mk}_{nk} \gamma_{mk}$. 
The MF ground state is that of the MF Hamiltonian \equa{eqn:HMF}, and could be written as
$\ket{\text{MF}}  = \prod_{vk} \left(U^{vk}_{vk}+\sum_{c} U^{vk}_{ck} c^{\dagger}_{ck} c_{vk} \right) \ket{\text{HF}}$
if the normal state $\ket{\text{HF}}$ is a semiconductor.
The gap equation at zero temperature reads
\begin{eqnarray}
\Delta_{vck} = \sum_{v'c'k'} V_{v'c'k',vck} \varphi_{v'c'k'},
\label{gap}
\end{eqnarray}
where $\varphi_{vck} \equiv  \langle c_{vk}^\dagger c_{ck} \rangle 
= \sum_{v_0} U^{v_0k*}_{vk} U^{v_0k}_{ck}$.
In the Bose-Einstein Condensation (BEC) limit where the exciton size is much smaller than the inter-exciton distance, $\varphi_{vck}$ may be interpreted as the excitonic wave function.
From \equa{SCF} and \equa{gap}, the quasiparticle energies, $U$-matrix and EI state can be calculated self-consistently.

We now present the BSE for the collective excitations.
The fluctuation Hamiltonian is
$H_{\text{fluc}} 
=- \sum_{vck,v'c'k'} V_{vck,v'c'k'} 
\left(
c^{\dagger}_{c'k'} c_{v'k'} - \varphi_{v'c'k'}^\ast 
\right)
\left( c^{\dagger}_{vk} c_{ck} 
- \varphi_{vck}
\right)$
which in the quasi-particle basis reads
\begin{widetext}
\begin{eqnarray}
\begin{split}
H_{\text{fluc}} 
=& - \sum_{vck,v'c'k'} V_{vck,v'c'k'} 
\left(\sum_{n'm'} U^{m'k'*}_{c'k'} U^{n'k'}_{v'k'} \gamma^{\dagger}_{m'k'} \gamma_{n'k'} - \varphi_{v'c'k'}^\ast 
\right)
\left( \sum_{nm} U^{nk*}_{vk} U^{mk}_{ck} \gamma^{\dagger}_{nk} \gamma_{mk}  
- \varphi_{vck}
\right),
\end{split}
\label{Hf}
\end{eqnarray}
\end{widetext}
With the interaction term $H_{\text{fluc}}$ turned on,
the ground state is modified from $\ket{\text{MF}}$ to the true ground state $\ket{0}$ which is a linear combination of $\ket{\text{MF}}$ and states with multi quasi $e$-$h$ pairs.
The (one pair) collective excited states on top of the true ground state 
may be written as
\begin{eqnarray}
\ket{S} = \sum_{vck} 
\left(
A^{S*}_{vck} \gamma^{\dagger}_{ck}\gamma_{vk}  - 
B^{S*}_{vck}\gamma^{\dagger}_{vk} \gamma_{ck}   
\right)
\ket{0}
\label{ES}
\end{eqnarray}
labeled by the integer $S$.
From the equations-of-motion (EOM) method~\cite{RMP1968,Fetter}, one obtains the BSE for the excited states in the matrix form (Appendix~\ref{sec:BSE}):
\begin{eqnarray}
\begin{split}
     \begin{bmatrix}
     K+\Pi^{AA} & \Pi^{AB}  \\
    - \Pi^{AB\ast} & -K-\Pi^{AA\ast}  \\
     \end{bmatrix}
     \begin{bmatrix}
     A^S  \\
     B^S  \\
     \end{bmatrix} = \Omega_S
     \begin{bmatrix}
     A^S  \\
     B^S  \\
     \end{bmatrix},
\end{split}
\label{bse}
\end{eqnarray}
where $\Omega_S$ is the excitation energy and $K_{vck,v'c'k'} = (E_{ck}-E_{vk}) \delta_{vck,v'c'k'}$ is the kinetic energy of free quasi $e$-$h$ pair.
The BSE scattering kernels are
\begin{widetext}
\begin{eqnarray}
\begin{split}
 \Pi^{AA}_{vck,v'c'k'} &=-\sum_{v_1c_1, v'_1c'_1}\left(V_{v_1c_1k,v'_1c'_1k'} U^{c'k'*}_{c'_1k'} U^{v'k'}_{v'_1k'} U^{vk*}_{v_1k} U^{ck}_{c_1k} + V_{v'_1c'_1k',v_1c_1k} U^{vk*}_{c_1k} U^{ck}_{v_1k} U^{c'k'*}_{v'_1k'} U^{v'k'}_{c'_1k'}\right),\\
 \Pi^{AB}_{vck,v'c'k'} &=-\sum_{v_1c_1, v'_1c'_1}\left(V_{v_1c_1k,v'_1c'_1k'} U^{ck}_{c_1k} U^{vk*}_{v_1k} U^{c'k'}_{v'_1k'} U^{v'k'*}_{c'_1k'} + V_{v'_1c'_1k',v_1c_1k} U^{c'k'}_{c'_1k'} U^{v'k'*}_{v'_1k'} U^{ck}_{v_1k} U^{vk*}_{c_1k} \right).
\end{split}
\label{kernel}
\end{eqnarray}
\end{widetext}
Because the BSE matrix is self-adjoint with respect to $\sigma_3=\mathrm{diag}(\hat{I}, -\hat{I})$ in the $A,B$ space, the eigen vectors $X_S=(A^S, B^S)^T$ could be made orthonormal to each other with the properly defined inner product: $X_S^\dagger \sigma_3 X_{S'} =\delta_{SS'}$.
One may also view \equa{bse} as the eigen equation for determining the canonical bosonic operators 
$a^\dagger_S=\sum_{vck} 
\left(
A^{S*}_{vck} \gamma^{\dagger}_{ck}\gamma_{vk}  - 
B^{S*}_{vck}\gamma^{\dagger}_{vk} \gamma_{ck}   
\right)$ 
for collective modes, see Appendix~\ref{sec:BSE}. 
The normalization condition is simply the requirement of the bosonic commutation relation: $[a_S,a^\dagger_{S'}]=\delta_{SS'}$.
The eigen energies $\Omega_S$, coming in  positive-negative pairs, are equivalent to those from the Green's function method~\cite{SZY2020,Murakami.2020, DX}.
In the limit of a normal semiconductor with zero excitonic order parameter, the $U$-matrix is identity and \equa{bse} reduces to the usual Tamm-Dancoff Approximation (TDA) of BSE for excitons.

These excited states may be viewed as the first excited states of the bosonic modes corresponding to order parameter fluctuations~\cite{Phase,BaSh,SZY2020}.
Its profile $\delta\Delta_{vck}$ may be defined as \cite{SI}: 
\begin{align}
\delta\Delta_{vck} &= \mel{0}{\sum_{v'c'k'} V_{v'c'k',vck} c^{\dagger}_{v'k'}c_{c'k'} }{S} 
+(S \leftrightarrow 0).
\label{dDelta}
\end{align}

The neglected two-body interactions such as $e$-$e$ and $h$-$h$ interactions can be treated with the standard MF approach in the same manner.
Their MF terms should be added to \equa{SCF} and their fluctuation terms give extra contributions to the BSE kernel in \equa{kernel}~\cite{SI}.
We also note that in \equa{eqn:H}, the interaction matrix elements are those screened by high energy degrees of freedom not included in the model.
However, for the calculation of excitons and excitonic orders, one may also include the screening effect from the degrees of freedom in \equa{eqn:H} itself~\cite{Zhang.2015}, which may be rationalized by integrating out the degrees of freedom at energies higher than the excitonic order one cares about.
In practice, because the excitonic condensation leads to change of the dielectric function compared to the normal state, the gap equation and the dielectric screening of the interaction should be solved self-consistently.

\begin{figure}
\centering
\includegraphics[width=8.0cm,clip=true]{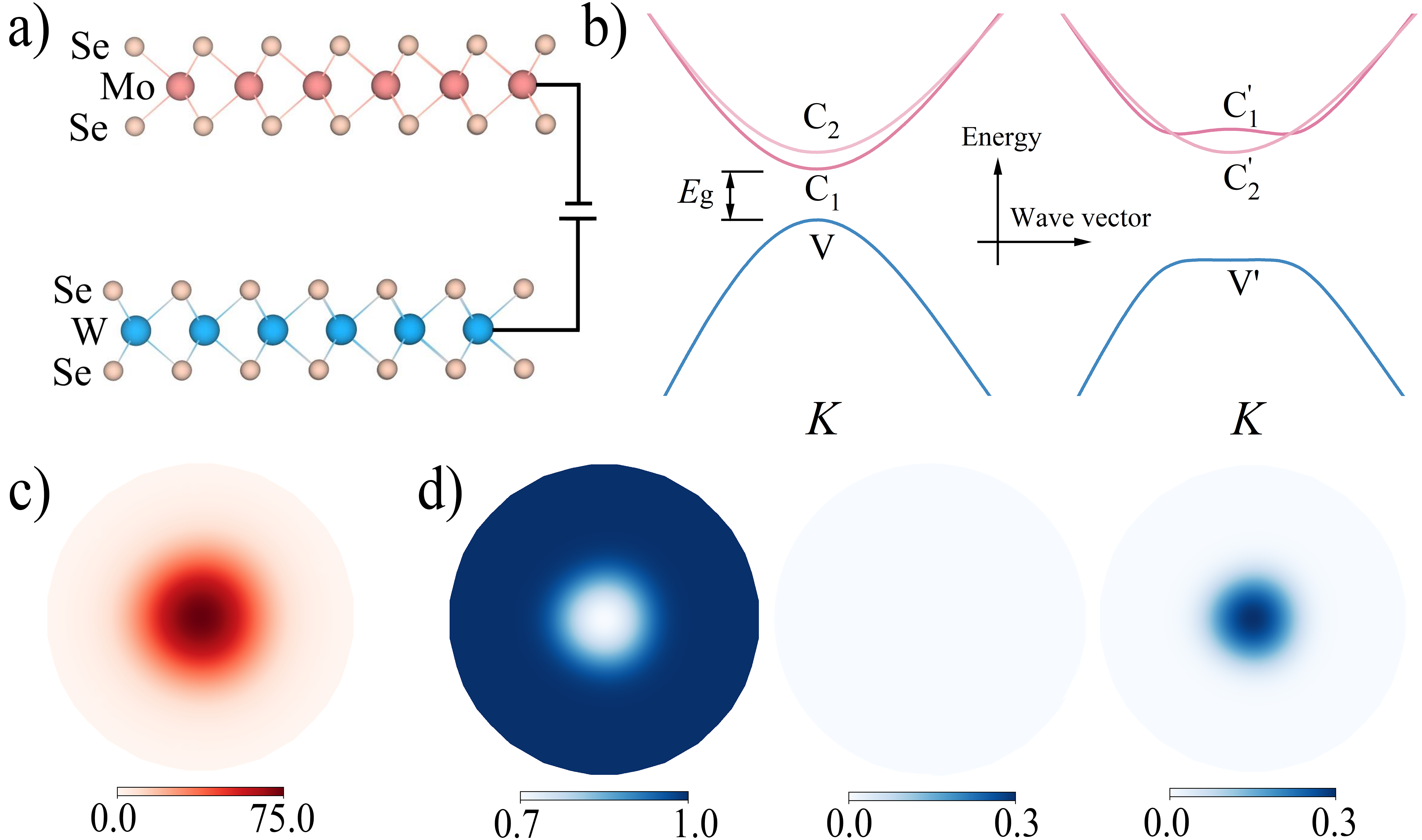}
\caption{(a) Atomic structure of the biased WSe$_2$-MoSe$_2$ bilayer. 
(b) Band structure of the normal state (left) and the EI state in the BEC regime (right) where $E_g=E_b/2$. 
(c) Plot of the order parameter $|\Delta_{vc_1k}|$ ($\unit{meV}$) in momentum space. (d) Plot of the $U$-matrix elements: $|U_{vvk}|^2$,$|U_{vc_2k}|^2$ and $|U_{vc_1k}|^2$ from left to right.}
\label{fig1}
\end{figure}

With Eqs.~1-9, we have described the general formalism suitable for first-principles calculation of the collective excitations in EIs without resorting to the Green's function technique. 
Next, we implement this method in existing software packages \cite{QE,QE2,BGW} and present results for the prototypical system, the biased WSe$_2$-MoSe$_2$ structure which is promising for realizing exciton condensation~\cite{Fogler2014, KFM2019,KFM2021,Wang2022,Du2023,KFM2025}.

As shown in \fig{fig1}(a), a vacuum layer equal to the thickness of monolayer hexagonal Boron Nitride (hBN) is inserted in between the $AB$-stacked WSe$_2$-MoSe$_2$ so that interlayer tunneling is neglected. 
For simplicity, we neglect dielectric screening from the environment including hBN but include that from the WSe$_2$-MoSe$_2$ bilayer itself~\cite{NatMater,Xuan2019}.
With no bias/gate voltage applied, the system is a semiconductor of type-II band alignment with the valence/conduction band residing on the WSe$_2$/MoSe$_2$ layer, respectively, and the $GW$ band gap is $E_g\approx2.0 \unit{eV}$.
The binding energy of interlayer excitons is calculated to be $E_{b}\approx 130 \unit{meV}$, not enough to overcome the band gap.
To reduce the band gap, one may apply an out-of-plane electric field using a gate voltage, or more practically, apply a chemical potential bias $\mu$ using electrical contacts shown in \fig{fig1}(a).
The effect of the latter is equivalently modeled as a shift of the conduction band ($\xi_{ck} = \epsilon_{ck} - \mu$) on  the MoSe$_2$ layer in \equa{eqn:H} so that the band gap is reduced by the same amount.
As the gap is reduced to be smaller than the binding energy, the excitonic order may develop.
Note that due to the absence of interlayer tunneling, the device we consider here is equivalent to an equilibrium system, although the bias combined with non-zero interlayer tunneling will drive the system into non-equilibrium steady states~\cite{SZY2024,Zeng.2024,Osterkorn.2025}.

Because the spin-orbital-coupling (SOC) leads to a splitting of the valence band by $\sim470 \unit{meV}$, while the splitting of the conduction band is $\sim20 \unit{meV}$, 
we include one valence band ($v$) from WSe$_2$ and two SOC-split conduction bands ($c_1$/$c_2$) from MoSe$_2$ in \equa{eqn:H}, as shown in the left of \fig{fig1}(b).
Moreover, we focus on the $K$-valley in the following since the $K$ and $K'$ valleys are degenerate in energy~\cite{SI}.
As a dense Brillouin zone (BZ) $k$-grid is essential to capture the $e$-$h$ interaction correctly, we adopt the nonuniform BZ sampling method \cite{nonuniform} using a spherical patch with radius 0.21 \AA$^{-1}$ around the $K/K'$-valley on a $210\times210$ $k$-grid, see Supplemental Material~\cite{SI,1986Louie,1987Louie,2000BSE,ONCV,ONCV2,PBE,Grimme,slab,Bogo1,Bogo2,Dong,Thouless1961} for computational details.

\begin{figure}
\centering
\includegraphics[width=8.0cm,clip=true]{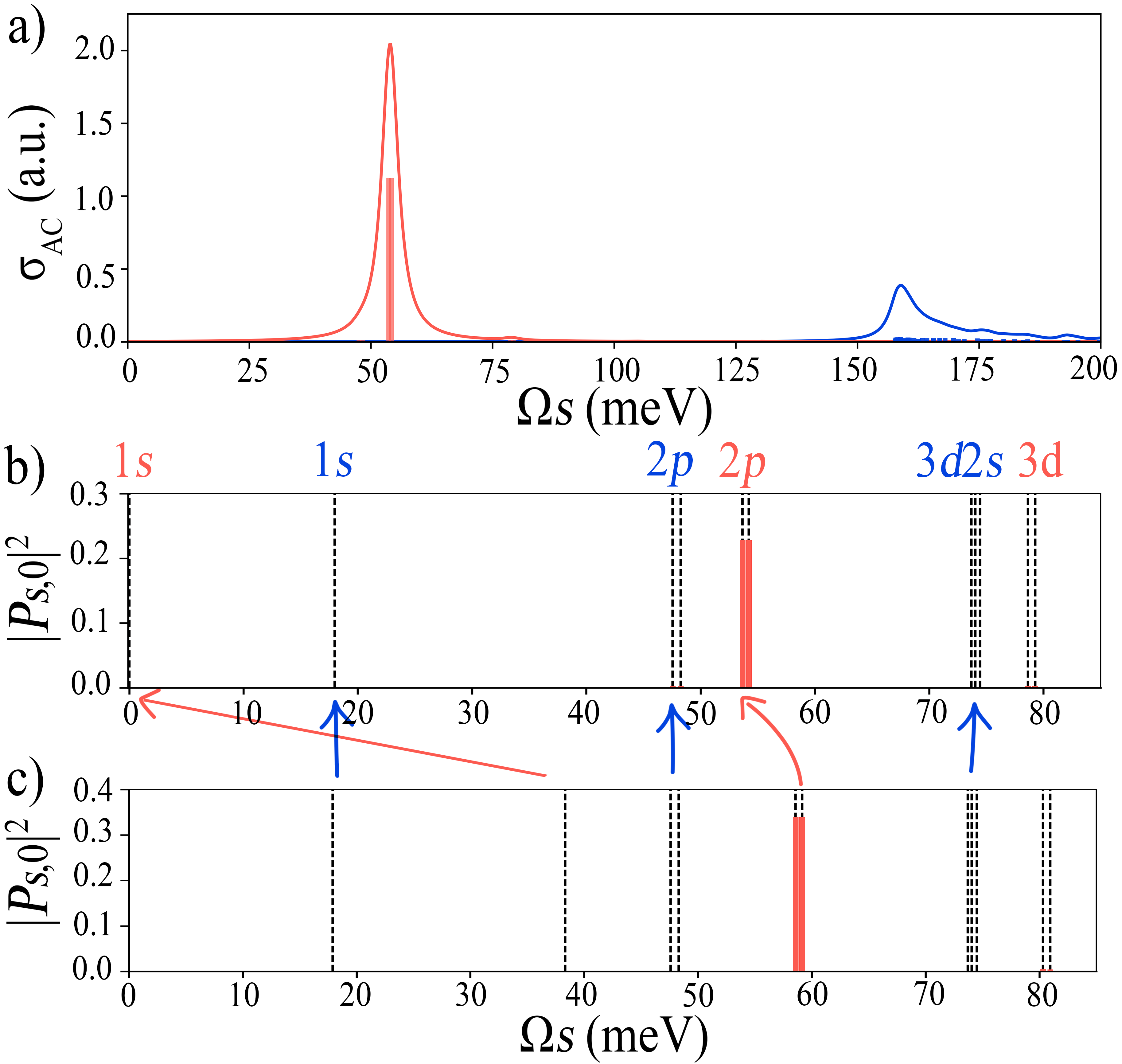}
\caption{(a) Optical conductivity along the armchair (AC) direction of the device in \fig{fig1}(a) in the BEC regime where $E_g=E_b/2$. Red and blue curves are calculated by BSE and IPA respectively with a Lorentzian broadening of $2 \unit{meV}$. 
(b) The excitation spectrum labeled by corresponding quantum numbers calculated from the BSE in \equa{bse}. Red and blue labels are for $c_1$- and $c_2$-types, respectively. 
(c) Same as (b) but calculated from TDA. The arrows between (b) and (c) connect the same labels.
}
\label{fig2}
\end{figure}

\fig{fig1}(b) shows the normal-state band dispersion $\xi_{nk}$ in the $K$-valley (left) and the resulting quasiparticle band $E_{nk}$ in the EI state from solving the gap equation~\eqref{gap} self-consistently , but with a static dielectric screening kept the same as that of the normal state.
The only inputs are $\xi_{nk}$ and $V_{vck,v'c'k'}$ which are from \textit{ab initio} calculations \cite{QE,QE2,BGW}.
We tuned the gap by the bias to $E_g=E_b/2$ so that the EI state is in the BEC regime. 
The originally lower-in-energy $c_1$-band is pushed up and the higher $c_2$-band is unchanged.
In the original band basis, the excitonic order parameter has 2-components: $(\Delta_{vc_1k},\Delta_{vc_2k})$.
From \fig{fig1}(c-d), we observe that an $s$-wave excitonic order parameter and non-zero $U$-matrix element hybridizing the valence band and the $c_1$ component are developed, while $\Delta_{vc_2k}$ and $U_{vc_2k}$ remain zero. 

\begin{figure}
\centering
\includegraphics[width=8.0cm,clip=true]{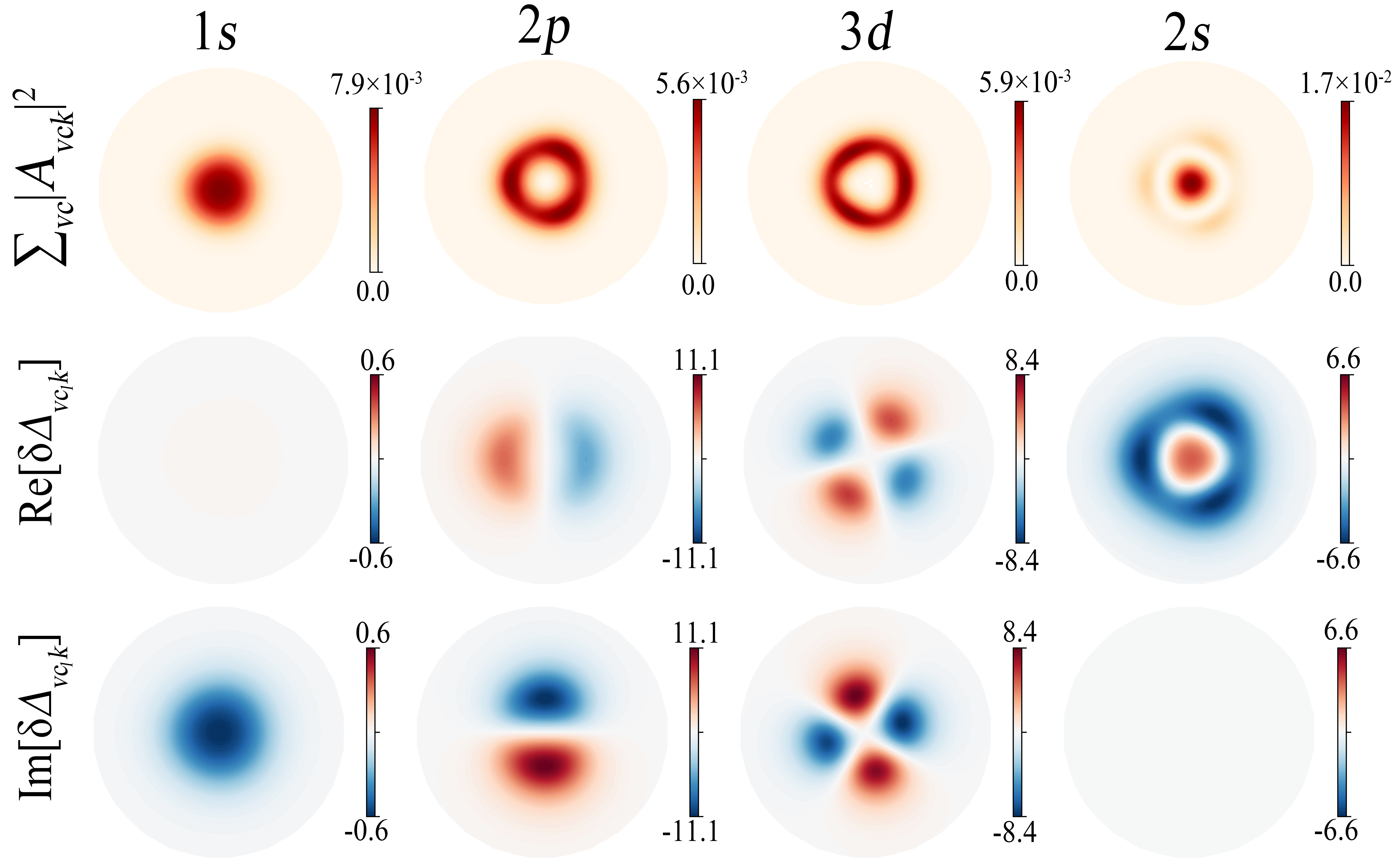}
\caption{
Plot of $\sum_{vc}|A_{vck}|^2$ and the real and imaginary parts of $\delta \Delta_{vc_1k}$ (meV) for a few $c_1$-type collective excitations on the momentum plane corresponding to those in \fig{fig2}.
}
\label{fig3}
\end{figure}

With the $U$-matrix from the MF calculation, we build the BSE matrix in \equa{bse} and solve it numerically for the collective excitations. 
The solutions can be categorized into two groups named the $c_1$-type (dominated by $A_{vc_1k}$ meaning excitation to the $c_1$ band) and $c_2$-type (dominated by $A_{vc_2k}$  meaning excitation to the $c_2$ band) shown by red and blue labels in \fig{fig2}b, respectively.
The linear optical conductivity can be calculated with the optical transition matrix elements\cite{SI}
\begin{eqnarray}
\begin{split}
\bra{S} \hat{p} \ket{0}= \sum_{vck} ( p'_{cvk} A^S_{vck} +p'_{vck} B^S_{vck}),
\end{split}
\label{pmatrix}
\end{eqnarray}
where
\begin{eqnarray}
\begin{split}
p'_{nmk} = \sum_{n'm'} p_{n'm'k} U^{nk*}_{n'k} U^{mk}_{m'k}.
\end{split}
\end{eqnarray}
\fig{fig2} shows the optical conductivity and the excitation spectrum.
In \fig{fig3} we plot the $k$-space distribution of $\sum_{vc}|A_{vck}|^2$ and the  profiles of order parameter fluctuations for $c_1$-type excitations. 
Because there is an arbitrary phase associated with the Bloch states \cite{Xuan2020},
the order parameter fluctuations in \fig{fig3} are plotted as $e^{-i\theta_k} \delta \Delta_{vck}$ with the momentum-dependent phase 
from the MF order parameter $\Delta_{vck} = e^{i\theta_k} |\Delta_{vck}|$, so that their relative phase is emphasized.
From \fig{fig3}, the angular momenta and radial quantum numbers can be identified for the excitations.

For the $c_1$-type excitations, there is a zero-energy solution that  corresponds to the phase-mode~\cite{Murakami.2020,SZY2020, Golez.2020}. 
It has a $s$-wave momentum dependence as shown in \fig{fig3} and the eigenvector fulfills $|A_{vck}|=|B_{vck}|$. 
Note that the phase difference between $A_{vck}$ and $B_{vck}$ is related to the gauge freedom in $U$-matrix.
It satisfies $A_{vck}=-B_{vck}$ for a gauge such that $U_{nvk}$ is real.
The $c_1$-type solutions with non-zero energies are the BaSh modes~\cite{SZY2020}.
All the BaSh modes with non-zero angular momentum  are (nearly) doubly degenerate while all the $s$-wave ones are non-degenerate.

In the normal state, the $v$ and $c_1$ bands have opposite spin and layer indices, meaning that there should be no interband optical transitions.
However, the new bands in the EI state are mixtures between them so that large interband optical matrix elements naturally exist~\cite{SI}. 
In the BEC case, these transitions could be viewed as the internal transitions of the excitons from the $1s$ bound state to higher bound states and scattering states.
From the approximate rotational symmetry of the bands, all the $p$-wave BaSh modes are optically active.
In \fig{fig2}, one observes that compared to the independent particle approximation (IPA), the interaction between the quasi-particles rearranges the  optical spectra weight from the continuous above-gap excitations 
to nearly a single strong peak at the $2p$-mode.
The optical activity of the subgap $2p$-mode may serve as a direct evidence of the presence of excitons in equilibrium.
Note that after the Hamiltonian of the electromagnetic field is included, the optically active BaSh modes will hybridize with photons to from BaSh polaritons which exhibit longitudinal-transverse splitting, similar to infrared phonons and bright excitons in solids. 
In three dimensions, this splitting shows up already at the zero wave vector limits of the modes. 
In the two-dimensional system we study, this effect shows up only at nonzero wavevectors, see Refs.~\cite{SZY2020, DX}.

The $c_2$-type excitations correspond to the fluctuation of $\delta\Delta_{vc_2k}$ (see Fig. S1 \cite{SI}), and can be classified by the angular quantum in similar ways.
From \fig{fig2}, one observes that all the $c_2$-type excitations are optically dark.
This is because the excitonic order $\Delta_{v c_1 k}$ does not mix the $c_2$ band with the other two bands so that the interband optical matrix elements $p'_{vc_2k}$ vanish.
The non-zero energy ($\sim20 \unit{meV}$) of the $1s$ mode originates from the SOC splitting of the two conduction bands, and will become another gapless mode in the limit of zero SOC splitting (see Fig. S2 \cite{SI}).

We note that for excitons in semiconductors, one may use the TDA where $\Pi^{AB}$ and $B_{vck}$ in \equa{bse} are neglected~\cite{2000BSE}.
However, in EIs, the full BSE is necessary to capture the collective modes because the particle-hole pair annihilation channel is an important portion of them. 
\fig{fig2}(b-c) show that compared to the result of TDA, the full BSE leads to substantial corrections for the $c_1$-type excitations. 
Nevertheless, TDA is a good approximation for $c_2$-type excitations~\cite{SI}.

\begin{figure}
\centering
\includegraphics[width=8.0cm,clip=true]{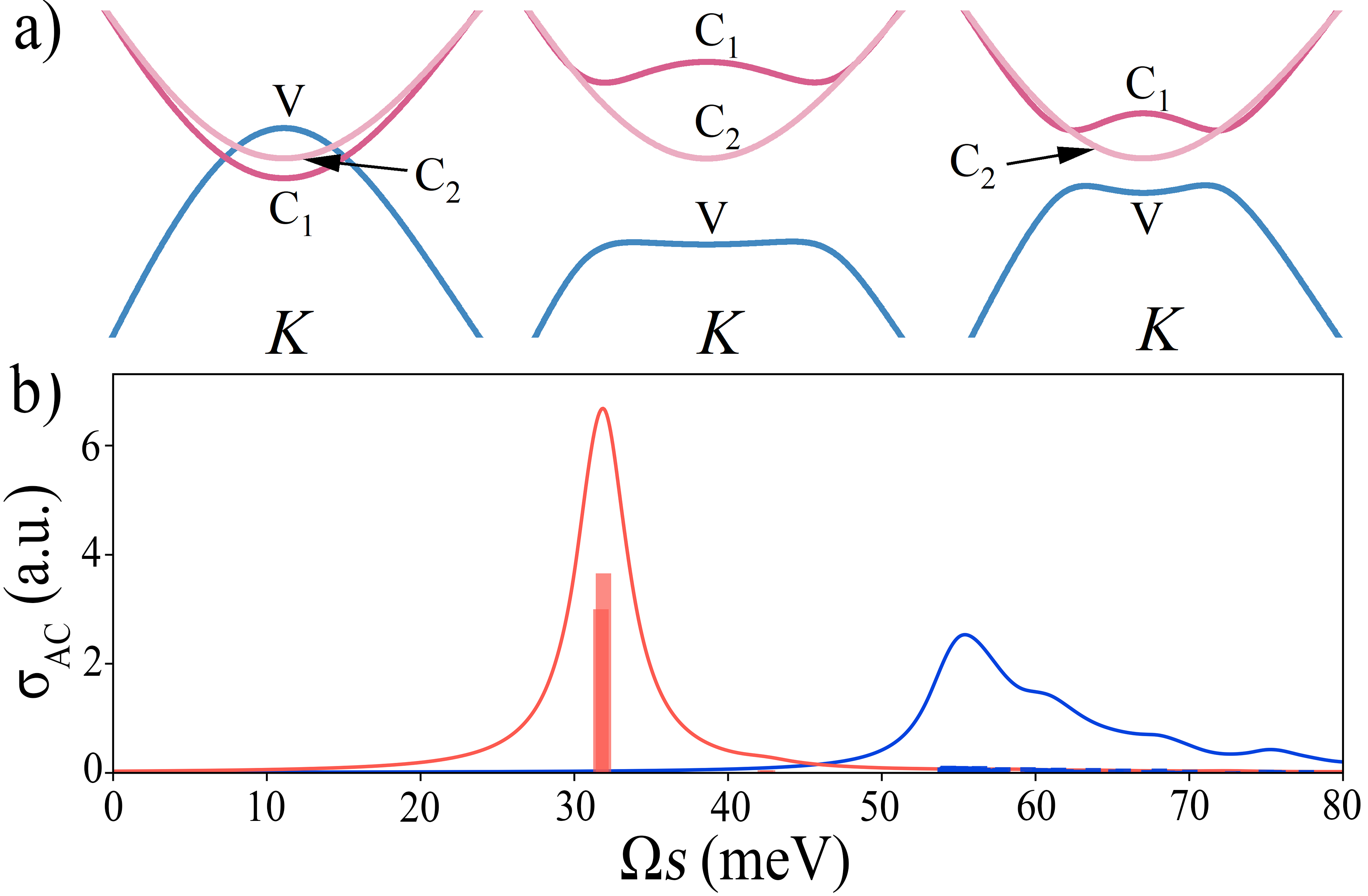}
\caption{(a) The BCS EI regime of the device in \fig{fig1}(a) where the gap is tuned to $E_g\approx-30\unit{meV}$.
Plotted are the band structures of the normal state (left) and the EI state without (middle) and with (right) updated dielectric screening. 
Note that the $c_2$-band remains unoccupied in this parameter range.
(b) Optical conductivity along the AC direction.
Red and blue curves are calculated from BSE and IPA respectively with a Lorentzian broadening of $2 \unit{meV}$.}
\label{fig4}
\end{figure}

Finally, in the dilute limit of the BEC case, the excitations are trivially understood as the internal transitions of a single exciton from the $1s$ bound state to higher states.
However, many-exciton effect starts to be important even slightly away from the dilute limit, which leads to dramatic shifts of the excitation energies and even their crossings, see Fig.~S3 ~\cite{SI}.
Therefore, the scheme described in Eqs.~2-9 is indispensable for obtaining the collective excitations in first-principles approaches to EIs beyond the single exciton picture.
This point is more obvious in the Bardeen–Cooper–Schrieffer (BCS) regime where the excitons are much larger than the inter-exciton distance so that the collective excitations depart qualitatively from the BEC case.
For example, the order parameter amplitude oscillation exists as the Higgs mode in the BCS weak coupling regime, while it and the phase oscillation become a conjugate pair of the same mode (i.e., the phase-mode) in the BEC case~\cite{SZY2020,Murakami.2020}.
In \fig{fig4}, we show a representative result in the BCS regime.
The optical conductivity shows that the $p$-wave BaSh mode is redshift significantly compared to that in the BEC regime.
As the exciton density is large, the extra screening from the condensate can be non-negligible.
Thus, we updated the dielectric screening using static random phase approximation with an approximated $k$-sampling~\cite{SI}. 
For more accurate results, a self-consistent calculation of the dielectric function with high numerical accuracy is needed in the future.

\begin{figure*}
\centering
\includegraphics[width=13cm,clip=true]{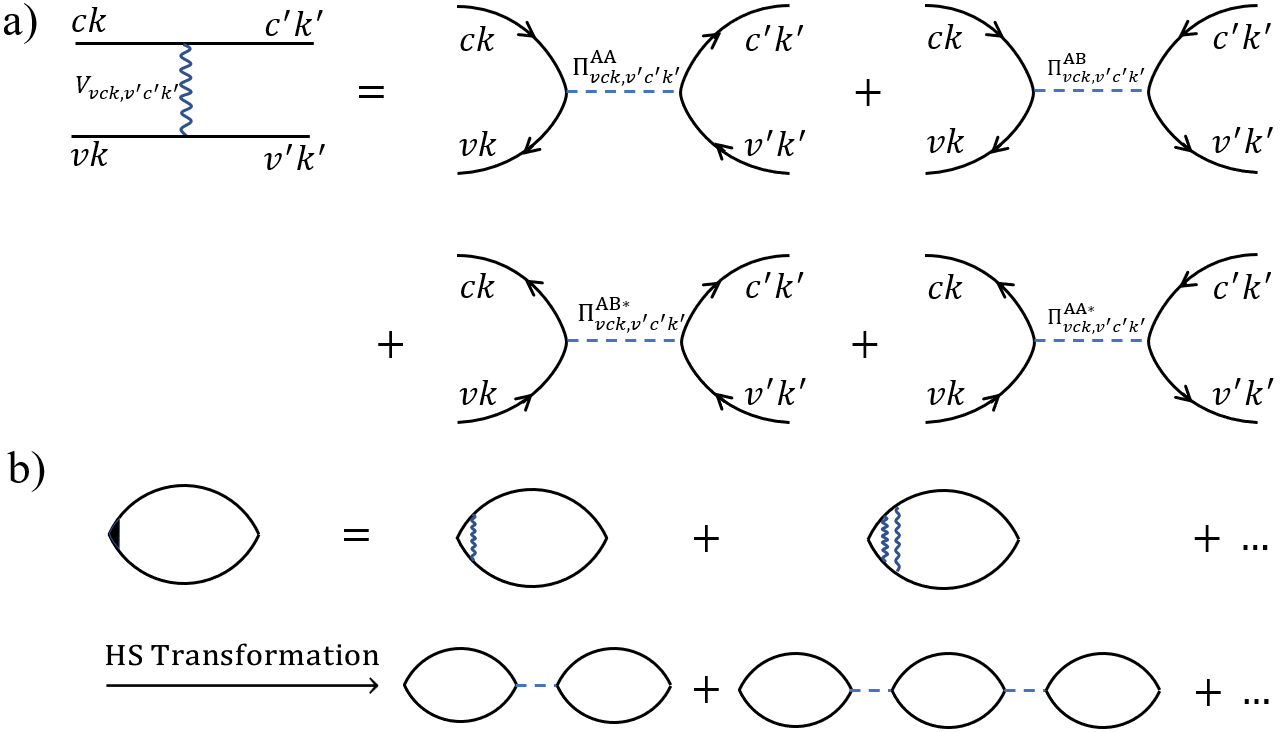}
\caption{
(a) Feynman diagrams for the $e$-$h$ propagator which yields the BSE. 
The two incoming legs may be viewed as the creation of a boson by $a^\dagger_S$,
so that this diagram is actually the self energy of the bosonic propagator.
Black solid lines are electronic propagators, which are corrected by the mean field if the system is in the  broken symmetry phase.
We use the blue wavy line to represent the interband interaction $V_{vck,v'c'k'}$ that includes both the direct and exchange interactions.
The blue dashed line represents the same interaction, but explicitly decomposed in the $e$-$h$ and anti-$e$-$h$ channels in the diagonalized band basis.
(b) Representation of collective modes in terms of bosonic propagators.
The first line is the bosonic propagator expressed in terms of the original interband interaction $V_{vck,v'c'k'}$.
The collective modes are contained in the vertex~\cite{PhysRevB.26.4883,BaSh}.
The second line are the equivalent diagrams after the Hubbard-Stratonovich transformation, where blue dashed lines are the propagators of the bosonic field that mediates the interaction $V_{vck,v'c'k'}$~\cite{Sun.2020_SC,SZY2020}.
}
\label{fig5}
\end{figure*}

In conclusion, we presented an \textit{ab initio} approach for computing the collective excitations in EIs.
It is also directly applicable to other systems such as charge/spin density waves (with no changes to Eqs.~1-9)~\cite{Gruner.1988_CDW, Gruner.1994_SDW, Sun.2024} and BCS superconductors (with a particle-hole transformation of the valence band operators in \equa{eqn:H})~\cite{Sun.2020_SC}.
After integration with existing first-principles software packages, we anticipate this method to enable \textit{ab initio} studies of excited-states in symmetry-broken quantum systems with state-of-the-art accuracy, thus offering concrete guidance for experiments.
However, we note that retardation effects of the electron-electron interaction arise in these systems due to either dynamical screening or mediation by phonons.
To further increase the accuracy of the first-principles approaches in the future, it is important to go beyond the instantaneous approximation to the interaction.
Finally, the BSE formalism is based on the picture of mean field plus Gaussian fluctuations, which works reasonably well for conventional EIs, superconductors and charge/spin density waves in the BCS weak coupling regime.
There the validity of the mean field approximation is known to be controlled by a small Ginzburg parameter~\cite{Ginzburg1961Ferroelectrics, Goldenfeld2018,Coleman,PRXSun}. 
Very close to the critical point, or for materials away from the BCS weak coupling regime, the methods addressing strong fluctuations~\cite{RevModPhys.47.773} need to be incorporated into \textit{ab initio} approaches in the future.

This work is funded by the National Science and Technology Major Project (Grants No. 2023ZD0120702),
Basic Research Program of Jiangsu (Grants No. BK20240395), 
the National Natural Science Foundation of China (Grants No. 12374291 and No. 12421004), 
the National Key Research and Development Program of China (2022YFA1204700), 
and Beijing Natural Science Foundation (Z240005).
Calculations were performed on Sugon HPC clusters equipped with HYGON X86 32-core processors (2.5 GHz) and at the Beijing Super Cloud Computing Center.
We thank Y. Murakami, D. Gole\v{z}, Y. Xu and C. Wang for helpful discussions.

\appendix
\section{}
\label{sec:BSE}

Here we derive the BSE with zero-momentum in \equa{bse} and \equa{kernel} using the EOM method in the quasiparticle representation.
EOM method \cite{RMP1968,Fetter} requires evaluation of commutation $[H,\gamma^{\dagger}_{ck}\gamma_{vk}]$ and $[H,\gamma^{\dagger}_{vk}\gamma_{ck}]$.
For the MF Hamiltonian, we have
\begin{eqnarray}
\begin{split}
\sum_{c'k'} [E_{c'k'} \gamma^{\dagger}_{c'k'}\gamma_{c'k'} ,\gamma^{\dagger}_{ck}\gamma_{vk}] &=  E_{ck} \gamma^{\dagger}_{ck}\gamma_{vk}, \\
\sum_{v'k'} [E_{v'k'} \gamma^{\dagger}_{v'k'}\gamma_{v'k'} ,\gamma^{\dagger}_{ck}\gamma_{vk}] &=  -E_{vk} \gamma^{\dagger}_{ck}\gamma_{vk}.
\label{eom2}
\end{split}
\end{eqnarray}
For the fluctuation Hamiltonian, we use \equa{Hf} to write out explicitly the particle-hole scattering matrix elements first, $\sum_{vck,v'c'k'} \Pi^+_{vck,v'c'k'} \gamma^{\dagger}_{c'k'} \gamma_{v'k'} \gamma^{\dagger}_{vk} \gamma_{ck}$ and $\sum_{vck,v'c'k'} \Pi^-_{vck,v'c'k'} \gamma^{\dagger}_{v'k'} \gamma_{c'k'} \gamma^{\dagger}_{vk} \gamma_{ck}$ where
\begin{widetext}
\begin{eqnarray}
\begin{split}
 \Pi^+_{vck,v'c'k'} &=-\sum_{v_1c_1, v_1'c_1'}\left( V_{v_1c_1k,v_1'c_1'k'} U^{c'k'*}_{c_1'k'} U^{v'k'}_{v_1'k'} U^{vk*}_{v_1k} U^{ck}_{c_1k} + V_{v_1'c_1'k',v_1c_1k} U^{vk*}_{c_1k} U^{ck}_{v_1k} U^{c'k'*}_{v_1'k'} U^{v'k'}_{c_1'k'} \right),  \\
 \Pi^-_{vck,v'c'k'} &=-\sum_{v_1c_1, v_1'c_1'}V_{v_1c_1k,v_1'c_1'k'} U^{ck}_{c_1k} U^{vk*}_{v_1k} U^{c'k'}_{v_1'k'} U^{v'k'*}_{c_1'k'} .
\end{split}
\end{eqnarray}
Next, taking the approximation $[\gamma^{\dagger}_{v'k'} \gamma_{c'k'} , \gamma^{\dagger}_{ck} \gamma_{vk}] \approx \delta_{vck,v'c'k'} $, we have
\begin{eqnarray}
\begin{split}
\sum_{vck,v'c'k'} \Pi^+_{vck,v'c'k'} [\gamma^{\dagger}_{c'k'}\gamma_{v'k'} \gamma^{\dagger}_{vk}\gamma_{ck} ,\gamma^{\dagger}_{ck_1}\gamma_{vk_1}] &= \sum_{v'c'k'} \Pi^+_{v_1c_1k_1,v'c'k'}  \gamma^{\dagger}_{c'k'}\gamma_{v'k'}
\,,
\\
\sum_{vck,v'c'k'} \Pi^-_{vck,v'c'k'} [\gamma^{\dagger}_{v'k'}\gamma_{c'k'} \gamma^{\dagger}_{vk}\gamma_{ck} ,\gamma^{\dagger}_{ck_1}\gamma_{vk_1}] &=  \sum_{v'c'k'} (\Pi^-_{v_1c_1k_1,v'c'k'} +\Pi^{-T}_{v_1c_1k_1,v'c'k'} ) \gamma^{\dagger}_{v'k'}\gamma_{c'k'} .
\label{eom4}
\end{split}
\end{eqnarray}
\end{widetext}

Finally, inserting the commutations of \equa{eom2} and \equa{eom4} in the EOM,
\begin{eqnarray}
\begin{split}
\mel{S}{[H,\gamma^{\dagger}_{ck}\gamma_{vk}]}{0}&=\Omega_S A^{S}_{vck},\\
\mel{S}{[H,\gamma^{\dagger}_{vk}\gamma_{ck}]}{0}&=\Omega_S B^{S}_{vck},
\end{split}
\end{eqnarray}
we obtain the BSE in the matrix form of \equa{bse},
where $A^S_{vck}=\mel{S}{\gamma^{\dagger}_{ck}\gamma_{vk}}{0}$ and $B^S_{vck}=\mel{S}{\gamma^{\dagger}_{vk}\gamma_{ck}}{0}$ which is normalized by $\sum_{vck} ( A^{S'}_{vck} A^{S*}_{vck} - B^{S’}_{vck}  B^{S*}_{vck}  ) = \delta_{SS'}$. 
For phase mode where $|A_{vck}|=|B_{vck}|$, we can normalize it with $\sum_{vck} |A_{vck}|^2 = 1$ \cite{Thouless1961}.
The BSE kernel in \equa{kernel} can be deduced from $\Pi^{AA}=\Pi^{+}$, $\Pi^{BB}=\Pi^{+*}$ and $\Pi^{AB}=\Pi^{-}+\Pi^{-T}$, $\Pi^{BA}=\Pi^{-*}+\Pi^{-*T}$.

The BSE can also be understood by the Feynman diagrams shown in \fig{fig5}(a).
The diagram on the left of \fig{fig5}(a) is viewed as the self energy for the propagator $G(\omega)$  of the zero center-of-mass-momentum boson
$a^\dagger_S=\sum_{vck} 
\left(
A^{S*}_{vck} \gamma^{\dagger}_{ck}\gamma_{vk}  - 
B^{S*}_{vck}\gamma^{\dagger}_{vk} \gamma_{ck}   
\right)$.
Expanded in the $(A, B)$ basis gives the four diagrams on the right, which is just the $\hat{\Pi}$ matrix in \equa{bse}.
Adding this self energy to the propagator of the free $e$-$h$ (and anti-$e$-$h$) pairs, one obtains 
\begin{align}
G(\omega)= 
\left[ 
\omega-      
\begin{pmatrix}
     K & 0 \\
    0 & -K  \\
\end{pmatrix}
-\hat{\Pi}
\right]^{-1}
\,.
\end{align}
Its pole condition gives the BSE in \equa{bse}.
Note that the off diagonal elements of $\hat{\Pi}$ mix $e$-$h$ and anti-$e$-$h$ pairs.


%

\clearpage

\setcounter{figure}{0} 
\setcounter{equation}{0} 
\renewcommand{\thefigure}{S\arabic{figure}}
\renewcommand{\theequation}{S\arabic{equation}}

\begin{widetext}

\begin{center}
\large

\textbf{Supplemental Material for: \textit{Ab initio} Approach to Collective Excitations in Excitonic Insulators}

\vspace{0.5cm}
\normalsize
Fengyuan Xuan,$^{1,*}$ Jiexi Song,$^1$ and Zhiyuan Sun$^{2,3,\dagger}$

\small
\textit{$^1$Suzhou Laboratory, Suzhou, 215123, China}

\textit{$^2$State Key Laboratory of Low-Dimensional Quantum Physics and Department of Physics, Tsinghua University, Beijing 100084, P. R. China}

\textit{$^3$Frontier Science Center for Quantum Information, Beijing 100084, P. R. China}

\vspace{3cm}

\textbf{I. SUPPLEMENTARY NOTES}

\end{center}

\textbf{A. Partition of the mean-field and fluctuation Hamiltonian}

\vspace{0.3cm}

In this section, we provide derivation for the partition of the mean-field (MF) and fluctuation Hamiltonian.
The starting point is the Hamiltonian with the electron-hole ($e$-$h$) interactions \cite{Kohn,Kel1965,Kel1968},
\begin{eqnarray}
     H = \sum_{ck} \xi_{ck} c^{\dagger}_{ck} c_{ck} + \sum_{vk} \xi_{vk} c^{\dagger}_{vk} c_{vk} - \sum_{vck,v'c'k',Q} V^Q_{vck,v'c'k'} c^{\dagger}_{c'k'+Q} c_{v'k'} c^{\dagger}_{vk} c_{ck+Q},
\end{eqnarray}
where $V^Q_{vck,v'c'k'}$ consists of direct and exchange interactions \cite{1986Louie,1987Louie,2000BSE},
\begin{eqnarray}
     V^Q_{vck,v'c'k'} = \mel{\psi_{c'k'+Q} \psi_{vk} }{W(r,r',\omega)}{\psi_{ck+Q} \psi_{v'k'}} - \mel{\psi_{c'k'+Q} \psi_{vk} }{\frac{1}{|r-r'|}}{ \psi_{v'k'} \psi_{ck+Q}}.
\label{Veh}
\end{eqnarray}
We next proceed with the zero-momentum case (Eq. 1 in the maintext) and omit the label $Q$. The formalism for $e$-$h$ pairing with non-zero center-of-mass momenta is similar.
We note that for $Q=0$ pairing, the $V^Q_{vck,v'c'k'}$ with finite-Q contribute to collective excitations with non-zero momenta.
The other two-body interaction terms such as electron-electron ($e$-$e$), hole-hole ($h$-$h$) interactions and intrinsic interband tunneling terms are neglected here and will be discussed in Section D.

By separating $c^{\dagger}_{vk} c_{ck}$ into its expectation value, $\varphi_{vck} \equiv  \langle c_{vk}^\dagger c_{ck} \rangle$, in the symmetry-broken ground state and a fluctuation term, the $e$-$h$ interaction is partitioned as
\begin{equation}
\begin{split}
	-\sum_{vck,v'c'k'} V_{vck,v'c'k'} c^{\dagger}_{c'k'} c_{v'k'} c^{\dagger}_{vk} c_{ck}  &= -\sum_{vck,v'c'k'} V_{vck,v'c'k'} \left(\varphi^*_{v'c'k'} +c^{\dagger}_{c'k'} c_{v'k'} - \varphi^*_{v'c'k'}\right)
 \left(\varphi_{vck} + c^{\dagger}_{vk} c_{ck}-\varphi_{vck} \right)\\
                 &= H_0 - \sum_{vck} \Delta^*_{vck} c^{\dagger}_{vk} c_{ck} - \sum_{vck}c^{\dagger}_{ck} c_{vk}\Delta_{vck} +H_{\text{fluc}},
\end{split}
\label{part}
\end{equation}
where $\Delta_{vck}=\sum_{v'c'k'} V_{v'c'k',vck} \varphi_{v'c'k'}$, $H_0$ and $H_{\text{fluc}}$ are,
\begin{eqnarray}
     H_0 = \sum_{vck,v'c'k'} V_{vck,v'c'k'}\varphi_{vck} \varphi^*_{v'c'k'},
\end{eqnarray}
\begin{eqnarray}
     H_{\text{fluc}} = -\sum_{vck,v'c'k'}V_{vck,v'c'k'}\left(c^{\dagger}_{c'k'} c_{v'k'} - \varphi^*_{v'c'k'} \right) \left(c^{\dagger}_{vk} c_{ck}-\varphi_{vck}\right).
\label{flucH1}
\end{eqnarray}
Therefore, the MF Hamiltonian reads

\begin{eqnarray}
\begin{split}
   H_{\text{MF}} &= H_0+ \sum_{ck} \xi_{ck} c^{\dagger}_{ck} c_{ck} + \sum_{vk} \xi_{vk} c^{\dagger}_{vk} c_{vk} - \sum_{vck} \Delta^*_{vck} c^{\dagger}_{vk} c_{ck}  - \sum_{vck} c^{\dagger}_{ck} c_{vk} \Delta_{vck} \\
     &= H_0 + \sum_k [... c^{\dagger}_{c_2k} c^{\dagger}_{c_1k} c^{\dagger}_{v_1k} c^{\dagger}_{v_2k} ...] h_k
     \begin{bmatrix} ...\\c_{c_2k}\\ c_{c_1k} \\ c_{v_1k}\\ c_{v_2k}\\... \end{bmatrix}\\
     &= H_0 + \sum_k [... c^{\dagger}_{c_2k} c^{\dagger}_{c_1k} c^{\dagger}_{v_1k} c^{\dagger}_{v_2k} ...]
     \begin{bmatrix}
     ... & ... & ... & ... & ... & ... \\ 
     ... &\xi_{c_2k} & 0 & -\Delta_{v_1c_2k} & -\Delta_{v_2c_2k} & ...\\ 
     ... &0 & \xi_{c_1k} & -\Delta_{v_1c_1k} & -\Delta_{v_2c_1k}& ... \\
     ...& -\Delta^*_{v_1c_2k} & -\Delta^*_{v_1c_1k} & \xi_{v_1k} & 0 &... \\ 
     ...& -\Delta^*_{v_2c_2k} & -\Delta^*_{v_2c_1k} & 0 &  \xi_{v_2k}&... \\
     ...& ... &... & ... & ... &  ... 
     \end{bmatrix}
     \begin{bmatrix} ...\\c_{c_2k}\\ c_{c_1k} \\ c_{v_1k}\\ c_{v_2k}\\... \end{bmatrix}\\
     &= H_0 + \sum_k [... c^{\dagger}_{c_2k} c^{\dagger}_{c_1k} c^{\dagger}_{v_1k} c^{\dagger}_{v_2k} ...] U
     \begin{bmatrix}
     ...& ... & ... & ... & ...  & ...\\
     ...& E_{c_2k} & 0 & 0 & 0  & ...\\ 
     ... &0 & E_{c_1k} & 0 & 0 & ... \\
     ... & 0 & 0 & E_{v_1k} & 0 & ...  \\ 
     ... & 0 & 0 & 0 & E_{v_2k} & ...\\ 
     ... & ... & ... & ... & ...  & ...
     \end{bmatrix}
     U^{\dagger} \begin{bmatrix} ...\\c_{c_2k}\\ c_{c_1k} \\ c_{v_1k}\\ c_{v_2k}\\... \end{bmatrix}\\
     &= H_0 + \sum_k [... \gamma^{\dagger}_{c_2k} \gamma^{\dagger}_{c_1k} \gamma^{\dagger}_{v_1k} \gamma^{\dagger}_{v_2k} ...] 
     \begin{bmatrix}
     ...& ... & ... & ... & ...  & ...\\
     ...& E_{c_2k} & 0 & 0 & 0  & ...\\ 
     ... &0 & E_{c_1k} & 0 & 0 & ... \\
     ... & 0 & 0 & E_{v_1k} & 0 & ...  \\ 
     ... & 0 & 0 & 0 & E_{v_2k} & ...\\ 
     ... & ... & ... & ... & ...  & ...
     \end{bmatrix}
     \begin{bmatrix} ...\\ \gamma_{c_2k}\\ \gamma_{c_1k} \\ \gamma_{v_1k}\\ \gamma_{v_2k} \\ ... \end{bmatrix} \\
&= H_0 + \sum_{ck} E_{ck} \gamma^{\dagger}_{ck} \gamma_{ck} + \sum_{vk} E_{vk} \gamma^{\dagger}_{vk} \gamma_{vk}.
\end{split}
\end{eqnarray}
In the above equation, the unitary transformation $U$-matrix with each column built by the eigenvectors of $h_k$ (Eq. 3 in the maintext) relates the $c$-operator for electrons with the $\gamma$-operator for quasiparticles \cite{Bogo1,Bogo2},
\begin{eqnarray}
\begin{split}
     \begin{bmatrix} ... \\ c_{c_2k} \\ c_{c_1k} \\ c_{v_1k} \\  c_{v_2k}\\ ...\end{bmatrix} = 
 \begin{bmatrix}
     ... & ... & ... & ... & ...& ...\\ 
     ... & U^{nk}_{c_2k} & ... & ... & U^{mk}_{c_2k} & ... \\ 
     ... & U^{nk}_{c_1k}& ... &... & U^{mk}_{c_1k} &... \\
     ... & U^{nk}_{v_1k} &... & ... & U^{mk}_{v_1k} &... \\
     ... & U^{nk}_{v_2k} & ... & ... & U^{mk}_{v_2k} & ... \\  
     ... & ... &... &... & ... &... \\
     \end{bmatrix}
\begin{bmatrix} ... \\ \gamma_{c_2k} \\ \gamma_{c_1k} \\\gamma_{v_1k} \\  \gamma_{v_2k}\\ ... \end{bmatrix}
= U \begin{bmatrix} ... \\ \gamma_{c_2k} \\ \gamma_{c_1k} \\\gamma_{v_1k} \\  \gamma_{v_2k}\\ ... \end{bmatrix}.
\end{split}
\end{eqnarray}
Using the $\gamma$-operator, the MF ground state can be constructed by
\begin{eqnarray}
     \ket{\text{MF}}=\prod_{vk} \gamma^{\dagger}_{vk} c_{vk}  \ket{\text{HF}} = \prod_{vk} \left(\sum_{n} \tilde{U}^{nk*}_{vk} c^{\dagger}_{nk} \right)c_{vk}  \ket{\text{HF}} 
     = \prod_{vk} \left(U^{vk}_{vk}+\sum_{c} U^{vk}_{ck} c^{\dagger}_{ck} c_{vk} \right) \ket{\text{HF}},
\end{eqnarray}
where $\tilde{U}^{nk}_{mk}=U^{mk*}_{nk}$, and $\ket{\text{HF}}$ is the conventional Hartree-Fock ground state for semiconductors. For the BCS regime, $\ket{\text{MF}}$ can be written in the same form if the valence and conduction bands are defined according to the Fermi surface. Note that we define $c/v$ according to the original band indices before applying the bias voltage in the maintext.

The self-consistent gap equation reads
\begin{equation}
\begin{split}
	\Delta_{vck} &= \mel{\text{MF}}{\sum_{v'c'k'} V_{v'c'k',vck} c^{\dagger}_{v'k'} c_{c'k'} }{\text{MF}} = \sum_{v'c'k'} V_{v'c'k',vck} \mel{\text{MF}}{c^{\dagger}_{v'k'} c_{c'k'}}{\text{MF}}= \sum_{v'c'k'} V_{v'c'k',vck} \varphi_{v'c'k'},
\end{split}
\label{gap}
\end{equation}
where $\varphi_{vck} \equiv  \langle c_{vk}^\dagger c_{ck} \rangle 
= \sum_{v_0} U^{v_0k*}_{vk} U^{v_0k}_{ck}$. 
Writing the $e$-$h$ interaction $- \sum_{vck,v'c'k'} V_{vck,v'c'k'} c^{\dagger}_{c'k'} c_{v'k'} c^{\dagger}_{vk} c_{ck}$ in the quasiparticle representation, we have
\begin{eqnarray}
\begin{split}
c^{\dagger}_{c'k'} c_{v'k'}= (U^{n'k'*}_{c'k'} \gamma^{\dagger}_{n'k'}) \times (U^{m'k'}_{v'k'} \gamma_{m'k'}), \quad
c^{\dagger}_{vk} c_{ck}= (U^{nk*}_{vk} \gamma^{\dagger}_{nk}) \times (U^{mk}_{ck} \gamma_{mk}).
\end{split}
\label{trans}
\end{eqnarray}
With the above, $H_{\text{fluc}}$ defined in Eq. \ref{part} and Eq. \ref{flucH1} is equivalent to the fluctuation Hamiltonian in the maintext,
and the derivation for the partition of the MF and fluctuation Hamiltonian is completed.

\vspace{1cm}

\textbf{B. Order parameter fluctuations and optical properties}

\vspace{0.3cm}

To evaluate fluctuations of the order parameter ($\delta\Delta_{vck}$) we define a perturbed ground state $\ket{0'} = \ket{0} + \alpha \ket{S}$, 
and $\delta\Delta_{vck}$ is expressed as
\begin{eqnarray}
\begin{split}
\delta\Delta_{vck} &= \lim_{\alpha \to 0} ( \mel{0'}{\sum_{v'c'k'} V_{v'c'k',vck} c^{\dagger}_{v'k'}c_{c'k'} }{0'} - \Delta_{vck} )/ \alpha\\
& = \mel{0}{\sum_{v'c'k'} V_{v'c'k',vck} c^{\dagger}_{v'k'}c_{c'k'} }{S} 
+ \mel{S}{\sum_{v'c'k'}V_{v'c'k',vck} c^{\dagger}_{v'k'}c_{c'k'} }{0},
\end{split}
\end{eqnarray}
which is the Eq. 9 in the maintext. 
Transforming from $c$-operators to $\gamma$-operators, we write down the expression to be used in first-principles calculations,
\begin{eqnarray}
\begin{split}
 \mel{S}{\sum_{v'c'k'} V_{v'c'k',vck} c^{\dagger}_{v'k'}c_{c'k'} }{0} & = \sum_{v'c'k'} V_{v'c'k',vck} \mel{S}{ U^{n'k'*}_{v'k'} \gamma^{\dagger}_{n'k'} U^{m'k'}_{c'k'}\gamma_{m'k'} }{0}\\
  & = \sum_{v'c'k'} V_{v'c'k',vck} U^{n'k'*}_{v'k'} U^{m'k'}_{c'k'} \mel{S}{  \gamma^{\dagger}_{n'k'} \gamma_{m'k'} }{0}\\
  & = \sum_{v''c'',v'c'k'} \left( V_{v'c'k',vck} U^{c''k'*}_{v'k'} U^{v''k'}_{c'k'} A^S_{v''c''k'} +V_{v'c'k',vck} U^{v''k'*}_{v'k'} U^{c''k'}_{c'k'} B^S_{v''c''k'}\right).
\end{split}
\end{eqnarray}
$\mel{0}{\sum_{v'c'k'} V_{v'c'k',vck} c^{\dagger}_{v'k'}c_{c'k'} }{S} $ can be obtained similarly.
Thus, the fluctuations of the order parameter can be readily calculated from first-principles.

For optical properties, the momentum operators are also transformed in the same manner,
\begin{eqnarray}
\begin{split}
P= \sum_{nmk} p_{nmk} c^{\dagger}_{nk} c_{mk} =\sum_{nmk} p'_{nmk} \gamma^{\dagger}_{nk} \gamma_{mk}, \quad p'_{nmk} = \sum_{n'm'} p_{n'm'k} U^{nk*}_{n'k} U^{mk}_{m'k}.
\end{split}
\end{eqnarray}
With Eq. 10 in the maintext, the optical conductivity is calculated using the Kubo formula,
\begin{eqnarray}
\begin{split}
\sigma_{\alpha\beta} &= \frac{i n e^2 \delta_{\alpha\beta} }{(\omega+i\eta)m} + \frac{i e^2}{(\omega+i\eta)  m^2 }   \sum_{S} \left( \frac{ \mel{0}{ P^{\alpha} }{S} \mel{S}{ P^{\beta} }{0} }{\hbar\omega -\Omega_S+i\eta}-\frac{ \mel{0}{P^{\beta}}{S} \mel{S}{P^{\alpha}}{0} }{\hbar\omega +\Omega_S+i\eta}  \right).
\end{split}
\end{eqnarray}

\vspace{1cm}

\textbf{C. 2-band form}

\vspace{0.3cm}

In the limit of 2-band configuration, the $U$-matrix can be written as \cite{Coleman},
\begin{eqnarray}
\begin{split}
    U=
         \begin{bmatrix}
     U^{ck}_{ck} & U^{vk}_{ck}  \\ 
     U^{ck}_{vk} & U^{vk}_{vk} \\ 
     \end{bmatrix}
     =
     \begin{bmatrix}
     u^*_k & v_k  \\ 
     -v^*_k & u_k \\ 
     \end{bmatrix},
\end{split}
\label{1v1cg}
\end{eqnarray}
and the BSE kernel matrix is reduced to
\begin{eqnarray}
\begin{split}
 \Pi^{AA}_{kk'} &=-V_{kk'} u^2_{k'} u^{*2}_{k} - V_{k'k} v^2_{k'} v^{*2}_{k}\\
 \Pi^{AB}_{kk'} &=V_{kk'} v^{*2}_{k'} u^{*2}_{k} + V_{k'k} u^{*2}_{k'} v^{*2}_{k}.
\end{split}
\end{eqnarray}
The interband momentum transition element $p'_{cvk} = (p_{cck}-p_{vvk})u_k v_k$ explains the brightness of collective modes of $c_1$-type excitations in excitonic insulators in the maintext ($p_{vck}$ is vanishing in the WSe$_2$-MoSe$_2$ bilayer due to the inserted vacuum of one layer hBN).
For the $c_2$-type, we have $\Pi^{AB}_{kk'} = 0$ as $v_k=U_{vc_2k} = 0$, which means TDA is good approximation.
For the $c_1$-type, we have finite $\Pi^{AB}_{kk'}$ as $v_k=U_{vc_1k}$ is nonvanishing which means that the $c_1$-type excitations need the full BSE. 
Using the gauge in this 2-band $U$-matrix (Eq. \ref{1v1cg}), we find that phase mode fulfills $A^P_{vck} = - B^P_{vck} e^{i\theta_k} $ where $e^{i\theta_k} $ is the phase of the order parameter $\Delta_{k}= e^{i\theta_k} |\Delta_k|$.

\vspace{1cm}

\textbf{D. Electron-electron and hole-hole interactions}

\vspace{0.3cm}

In general, the total many-body Hamiltonian reads \cite{Fetter}
\begin{eqnarray}
\begin{split}
     H_{tot} & = \sum_{ck} \xi_{ck} c^{\dagger}_{ck} c_{ck} + \sum_{vk} \xi_{vk} c^{\dagger}_{vk} c_{vk} +\frac{1}{2}\sum_{\alpha\beta\alpha'\beta'} V_{\alpha\beta\alpha'\beta'} N( c^{\dagger}_{\alpha}c^{\dagger}_{\beta}  c_{\beta'}  c_{\alpha'}),
\end{split}
\label{Htot}
\end{eqnarray}
where $N$ stands for normal-ordering with respect to the particle-hole representation of $c$-operators.
The Hamiltonian in the Eq. 1 of maintext keeps only the $e$-$h$ interaction in the two-body term.
In principle, all other two-body terms can be treated in the same way as Eq. \ref{trans}: 
first transform into quasiparticle representation by $c_{nk} = \sum_m U^{mk}_{nk} \gamma_{mk}$, 
then make them normal-ordered with respect to the $\gamma$-operator and it will end up with a one-body term and a two-body term in the quasiparticle representation.
To illustrate, we use a configuration with only one valence and one conduction band ($1V1C$) as presented in the Section C of this Supplementary Material. 
Apart from the $e$-$h$ interaction, we add the $e$-$e$ and $h$-$h$ direct interactions to the Hamiltonian,
\begin{eqnarray}
\begin{split}
     H' = &\sum_{k} \xi_{ck} c^{\dagger}_{ck} c_{ck} + \sum_{k} \xi_{vk} c^{\dagger}_{vk} c_{vk} - \sum_{kk'} V^{eh}_{kk'} c^{\dagger}_{ck'} c_{vk'} c^{\dagger}_{vk} c_{ck} 
+ \sum_{kk'} V^{ee}_{kk'} c^{\dagger}_{ck'} c_{ck'} c^{\dagger}_{ck} c_{ck} + \sum_{kk'} V^{hh}_{kk'} c_{vk'} c^{\dagger}_{vk'} c_{vk} c^{\dagger}_{vk} .
\end{split}
\end{eqnarray}
The sign in front of each interaction denotes the nature of attraction/repulsion.
The $h_k$ and $U$-matrix in this $1V1C$-configuration are written as
\begin{eqnarray}
\begin{split}
    h_k=
     \begin{bmatrix}
     \xi_{ck}+\xi^{ee}_{ck} & -\Delta_k  \\ 
     -\Delta^*_k & \xi_{vk}-\xi^{hh}_{vk} \\ 
     \end{bmatrix}, \quad
    U=
     \begin{bmatrix}
     u^*_k & v_k  \\ 
     -v^*_k & u_k \\ 
     \end{bmatrix},
\end{split}
\end{eqnarray}
where the extra one-body terms, $\xi^{ee}_{ck}$ and $\xi^{hh}_{vk}$, in $h_k$ come from the repulsive $e$-$e$ and $h$-$h$ interactions,
\begin{eqnarray}
\begin{split}
 \sum_{kk'} V^{ee}_{kk'} c^{\dagger}_{ck'} c_{ck'} c^{\dagger}_{ck} c_{ck} &= \sum_{kk'} V^{ee}_{kk'} (|v_{k'}|^2+c^{\dagger}_{ck'} c_{ck'} -|v_{k'}|^2 ) (|v_k|^2+c^{\dagger}_{ck} c_{ck} - |v_k|^2)\\
 &= H^{ee}_0  + \sum_{kk'} (V^{ee}_{kk'} +V^{ee}_{k'k}) |v_{k'}|^2 c^{\dagger}_{ck} c_{ck} + H^{ee}_{\text{fluc}} \\
 &= H^{ee}_0  + \sum_{kk'} \xi^{ee}_{ck} c^{\dagger}_{ck} c_{ck} + H^{ee}_{\text{fluc}}.
\end{split}
\end{eqnarray}
In the above equation, the two-body fluctuation term due to the $e$-$e$ interaction reads
\begin{eqnarray}
\begin{split}
 H^{ee}_{\text{fluc}}=&\sum_{kk'}  u_{k'}v_{k'}  u^*_kv^*_k (V^{ee}_{kk'} +V^{ee}_{k'k} ) \gamma^{\dagger}_{ck'} \gamma_{vk'}  \gamma^{\dagger}_{vk} \gamma_{ck}\\
&+\sum_{kk'}  u_{k'}v_{k'}  u_kv_k V^{ee}_{kk'}  \gamma^{\dagger}_{ck'} \gamma_{vk'}  \gamma^{\dagger}_{ck} \gamma_{vk}
+\sum_{kk'}  u^*_{k'}v^*_{k'}  u^*_kv^*_k V^{ee}_{kk'} \gamma^{\dagger}_{vk'} \gamma_{ck'}  \gamma^{\dagger}_{vk} \gamma_{ck}.
\end{split}
\end{eqnarray}
Similarly, we get $\xi^{hh}_{vk}$ and $H^{hh}_{\text{fluc}}$. 
Thus, the extra one-body terms of $\xi^{ee}_{ck}$ and $\xi^{hh}_{vk}$ modify the MF calculation through $h_k$, while the extra two-body terms enter the BSE kernel,
\begin{eqnarray}
\begin{split}
 \Pi^{AA}_{kk'} &=-V_{kk'} u^2_{k'} u^{*2}_{k} - V_{k'k} v^2_{k'} v^{*2}_{k} + u_{k'}v_{k'}  u^*_kv^*_k (V^{ee}_{kk'} +V^{ee}_{k'k} +V^{hh}_{kk'} +V^{hh}_{k'k}) \\
 \Pi^{AB}_{kk'} &=V_{kk'} v^{*2}_{k'} u^{*2}_{k} + V_{k'k} u^{*2}_{k'} v^{*2}_{k} + u^*_{k'}v^*_{k'}  u^*_kv^*_k (V^{ee}_{kk'} +V^{ee}_{k'k} +V^{hh}_{kk'} +V^{hh}_{k'k}).
\end{split}
\end{eqnarray}
Therefore, one could calculate all the many-body effects from first-principles as discussed in the maintext. 
For the numerical demonstration of WSe$_2$-MoSe$_2$ bilayer structure, the repulsive effect from $e$-$e$ and $h$-$h$ interactions can be compensated by the bias chemical potential $\mu$.
And the intrinsic interband tunneling effect from the two-body term in Eq. \ref{Htot}, containing one conduction (valence) and three valence (conduction) bands, can be neglected as the wave function overlap between the two layers is vanishing due to the inserted vacuum of one layer hBN.
So we take the simple Hamiltonian with only $e$-$h$ interactions in the numerics of maintext.
For bulk excitonic insulator materials, other many-body interactions can be important for accurate \textit{ab initio} calculations.

\vspace{1cm}

\textbf{E. Dielectric screening from exciton condensate}

\vspace{0.3cm}

In this section, we discuss the extra screening effect in the presence of exciton condensation.
The major effect of the $e$-$h$ interaction in Eq. \ref{Veh} is the direct term that depends on the screened Coulomb potential $W(r,r',\omega)$.
In reciprocal space, the screened Coulomb potential is determined by the dielectric matrix \cite{1986Louie,1987Louie},
\begin{eqnarray}
W_{GG'}(q,\omega) =\frac{1}{(2\pi)^6}\int e^{-i(q+G)r} W(r,r',\omega) e^{i(q+G')r'} drdr'=\epsilon^{-1}_{GG'}(q,\omega) v_{q+G},
\end{eqnarray}
where $v_{q+G}  = 4\pi/|q+G|^2$ is the Coulomb potential. 
Using the Random Phase Approximation, the dielectric matrix is calculated from the non-interacting polarizability matrix \cite{BGW}
\begin{eqnarray}
\epsilon_{GG'}(q,\omega) = \delta_{GG'} - v_{q+G} \chi_{GG'}(q,\omega),
\end{eqnarray}
We note that in principle one should calculate frequency-dependent dielectric function, which means that the gap equation and BSE should be solved self-consistently with updated $\epsilon_{GG'}(q,\omega)$.
In the numerical demonstration, we neglect the dynamical screening effect\cite{Dong} and the static polarizability matrix is calculated by \cite{1987Louie},
\begin{eqnarray}
\begin{split}
\chi_{GG'}(q,0) = \chi_{GG'}(q) = \frac{2}{\Omega} \sum_{vck} \frac{ \mel{\psi_{vk}}{ e^{-i(q+G)r}}{\psi_{ck+q}} \mel{\psi_{ck+q}}{ e^{i(q+G')r'}}{\psi_{vk}} }{E_{vk}-E_{ck+q}}.
\end{split}
\end{eqnarray}
For the normal Kohn-Sham state in WSe$_2$-MoSe$_2$, the wave function overlap between the two layers can be neglected and the total polarizability is the sum of the individual layer \cite{Xuan2019}.
As the excitonic order parameter is developed, the top valence and top conduction band on the two layers starts to mix into the quasiparticle wave function $\psi^{B}_{mk} = \sum_n U^{mk*}_{nk}\psi_{nk}$.
Therefore, after obtaining the EI state, we need to calculate the extra dielectric screening from the exciton condensate.
Here, due to limitation of computational resource, we approximate $\chi_{GG'}(q) \approx \chi^{\text{BL}}_{GG'}(q) + \chi^{\text{EI}}_{GG'}(q)$, 
where $\chi^{\text{BL}}_{GG'}(q)$ is the original polarizability of the bilayer and $ \chi^{\text{EI}}_{GG'}(q)$ is evaluated using only the quasiparticle transition at $K$ with an effective number of condensate region $k$-points on the $210\times210\times1$ $k$-grid,
\begin{eqnarray}
\begin{split}
\chi^{\text{EI}}_{GG'}(q) =  \frac{2}{\Omega}  \bar{N}_{K}\frac{ \mel{\psi^B_{vK}}{ e^{-i(q+G)r}}{\psi^B_{cK+q}} \mel{\psi^B_{cK+q}}{ e^{i(q+G')r'}}{\psi^B_{vK}} }{E_{vK}-E_{cK+q}},
\end{split}
\end{eqnarray}
where $\bar{N}_K = \sum_k |U_{vc_1k}|^2/|U_{vc_1K}|^2 $.
As the exciton density in the BCS regime is much larger than the BEC regime, 
we update the dielectric screening from $\chi^{\text{EI}}$ for the BCS regime as stated in the maintext.
We note that a convergent dielectric screening from the condensed excitons with better calculation accuracy can be potentially important in the BCS regime.

\vspace{1cm}

\textbf{F. Computational details}

\vspace{0.3cm}

The density functional theory (DFT) and many-body perturbation theory (MBPT) calculations in this work are performed within the QuantumESPRESSO \cite{QE,QE2} and BerkeleyGW~\cite{BGW} packages.
Optimized norm-conserving Vanderbilt pseudopotentials \cite{ONCV} with a kinetic energy cut-off of $60$ Ry and Perdew-Becke-Ernzerhof approximation of the exchange correlation functional \cite{PBE} are used.
The Brillouin zone (BZ) sampling $k$-grid for the DFT self-consistent calculation is $36\times36\times1$.
The $GW$ band gap without bias voltage is calculated using $12\times12\times1$ $q/k$-grid and $8$ Ry $G$-vector cutoff.
The DFT wave functions on the BZ-patch with radius of $0.21$ \AA$^{-1}$ from $K/K'$ on a dense $k$-grid of $210\times210\times1$ are prepared for the calculation of $e$-$h$ interaction kernel $V_{vck,v'c'k'}$.
In the Eq. 1 of the maintext, $\xi_{ck} = \epsilon_{ck} + \mu$ where $\epsilon_{nk}$ is the Kohn-Sham energies and the bias voltage $\mu$ is treated by scissor-operation.
A fine $q$-grid is generated by $q=k-k'$ where $k/k'$ is on the dense $k$-grid BZ-patch.
The dielectric matrix on this $q$-grid is calculated using $6\times6\times1$ $k$-grid sampling, $8$ Ry $G$-vector cutoff and 500 empty bands.
The slab coulomb truncation~\cite{slab} is adopted and no symmetry is used to reduce the $k$-grid throughout the calculation of dielectric function and the $e$-$h$ interaction kernel. 
After we get $\xi_{ck}$ and $V_{vck,v'c'k'}$ from standard DFT and MBPT calculations using QuantumESPRESSO and BerkeleyGW, 
the MF calculation for EI ground state includes diagonalization of $h_k$ and calculation of $\Delta_{vck}$ using Eq. \ref{gap} self-consistently.
Then we construct the BSE and solve it numerically for the many-body excitation energies and excited states.
The fluctuation of order parameter for each collective mode is calculated using Eq. 9 in the maintext.
We also performed calculation for the $1V1C$-configuration without spin-orbital-coupling (SOC) including both $K$ and $K'$ valley (see Fig.~\ref{KKp}).
$K$ and $K'$ valley are degenerate so that we focus on $K$ valley for the $1V2C$-configuration in the maintext.
Band convergence result in Fig.~\ref{band} further shows that higher energy conduction bands do not affect the excitation spectrum in the EI state.

\clearpage

\begin{center}
\textbf{II. SUPPLEMENTARY FIGURES}
\end{center}

\begin{figure*}[h!]
\centering
\includegraphics[width=16.0cm,clip=true]{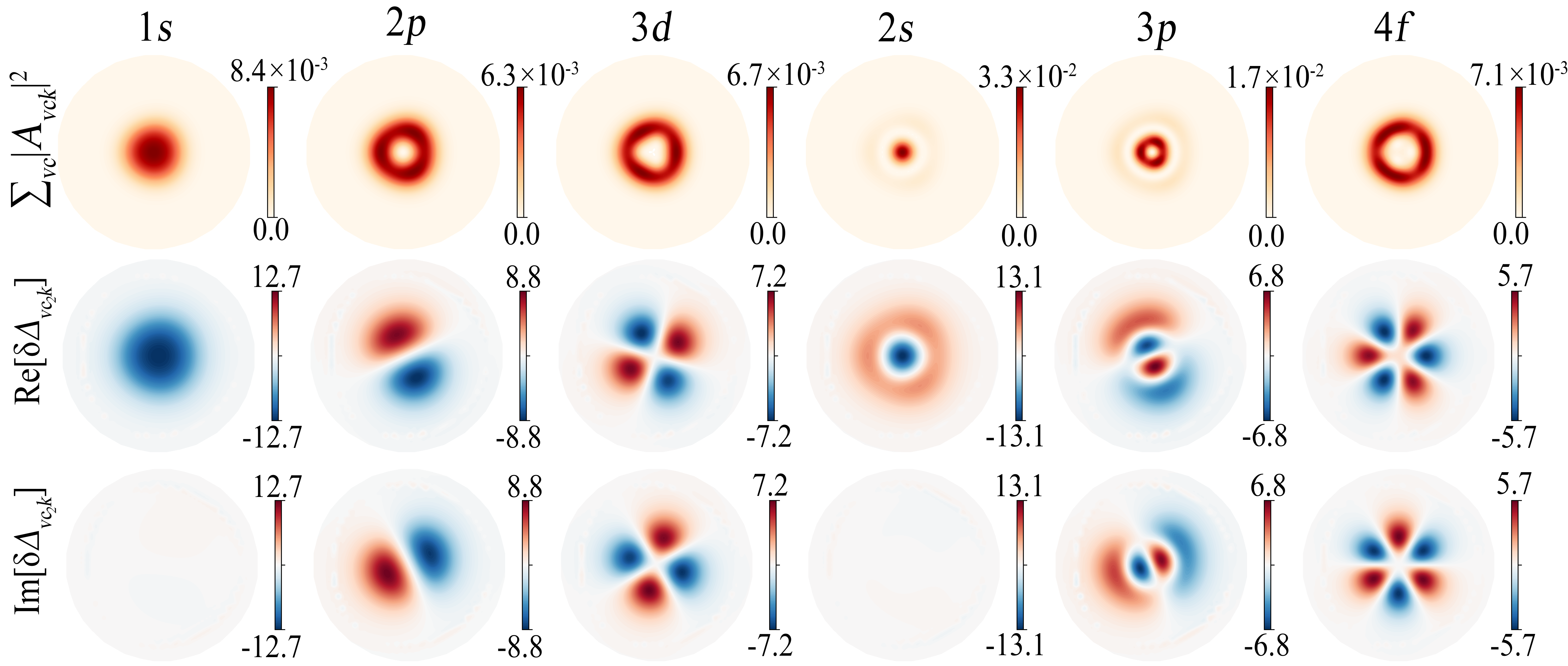}
\caption{Plot of $\sum_{vc}|A_{vck}|^2$ and the real and imaginary parts of $\delta \Delta_{vc_2k}$ (meV) for a few $c_2$-type collective excitations on the momentum plane corresponding to the blue labels in Fig. 2(b) of the maintext.}
\label{c2}
\end{figure*}
\begin{figure*}[h!]
\centering
\includegraphics[width=15.0cm,clip=true]{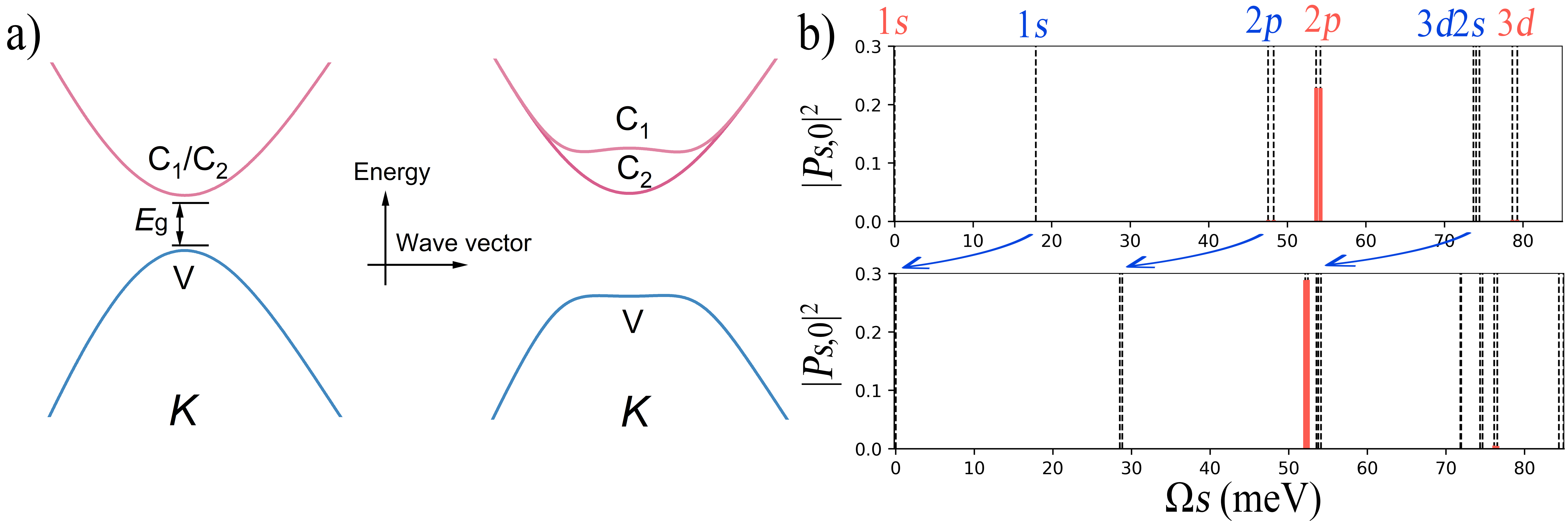}
\caption{(a) Band structure of normal state and EI state for $1V2C$-configuration with artificially degenerate conduction bands. 
(b) Excitation spectrum for $1V2C$-configuration with calculated SOC-splitting (upper) and artificially degenerate (lower) conduction bands. 
Red and blue labels on top refers to $c_1$- and $c_2$-type excitations, respectively. 
The $c_2$-type excitations are redshift by the SOC splitting indicated by the blue arrows.
The $1s$ $c_2$-type excitation becomes another gapless mode in the limit of zero SOC splitting as discussed in the maintext.}
\label{soc}
\end{figure*}
\begin{figure*}[h!]
\centering
\includegraphics[width=12.0cm,clip=true]{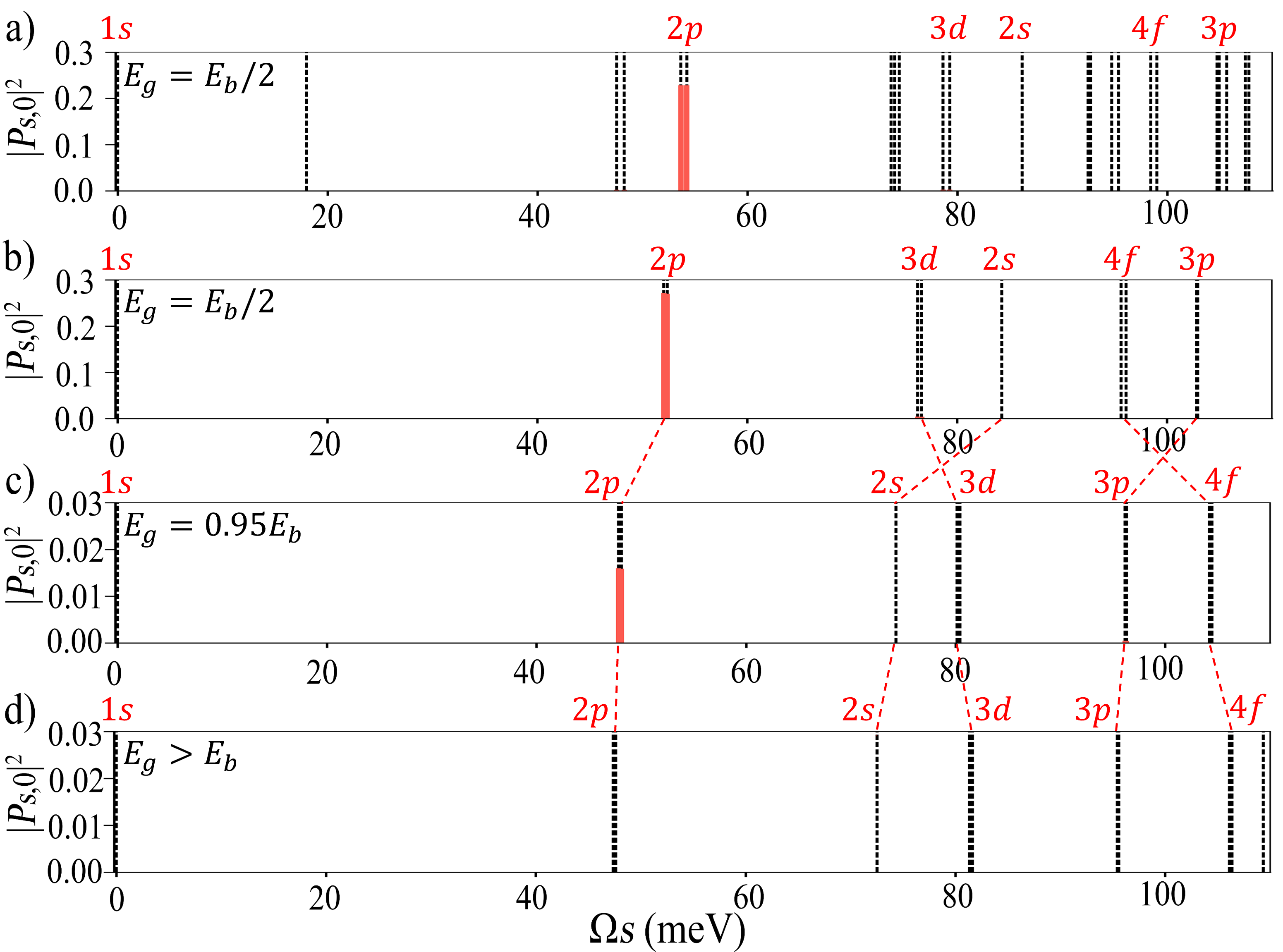}
\caption{The excitation spectrum labeled by corresponding quantum numbers calculated from the BSE. (a) $1V2C$-configuration results reproduced from Fig. 2(b) in the maintext. Only $c_2$-type excitations are labeled. (b-d) $1V1C$-configuration results with different $E_g$. For $E_g>E_b$, the excitation energies $\Omega_S$ are subtracted by the $1s$-state energy. (b-d) are computed without SOC effect.}
\label{vc}
\end{figure*}
\begin{figure*}[h!]
\centering
\includegraphics[width=15.0cm,clip=true]{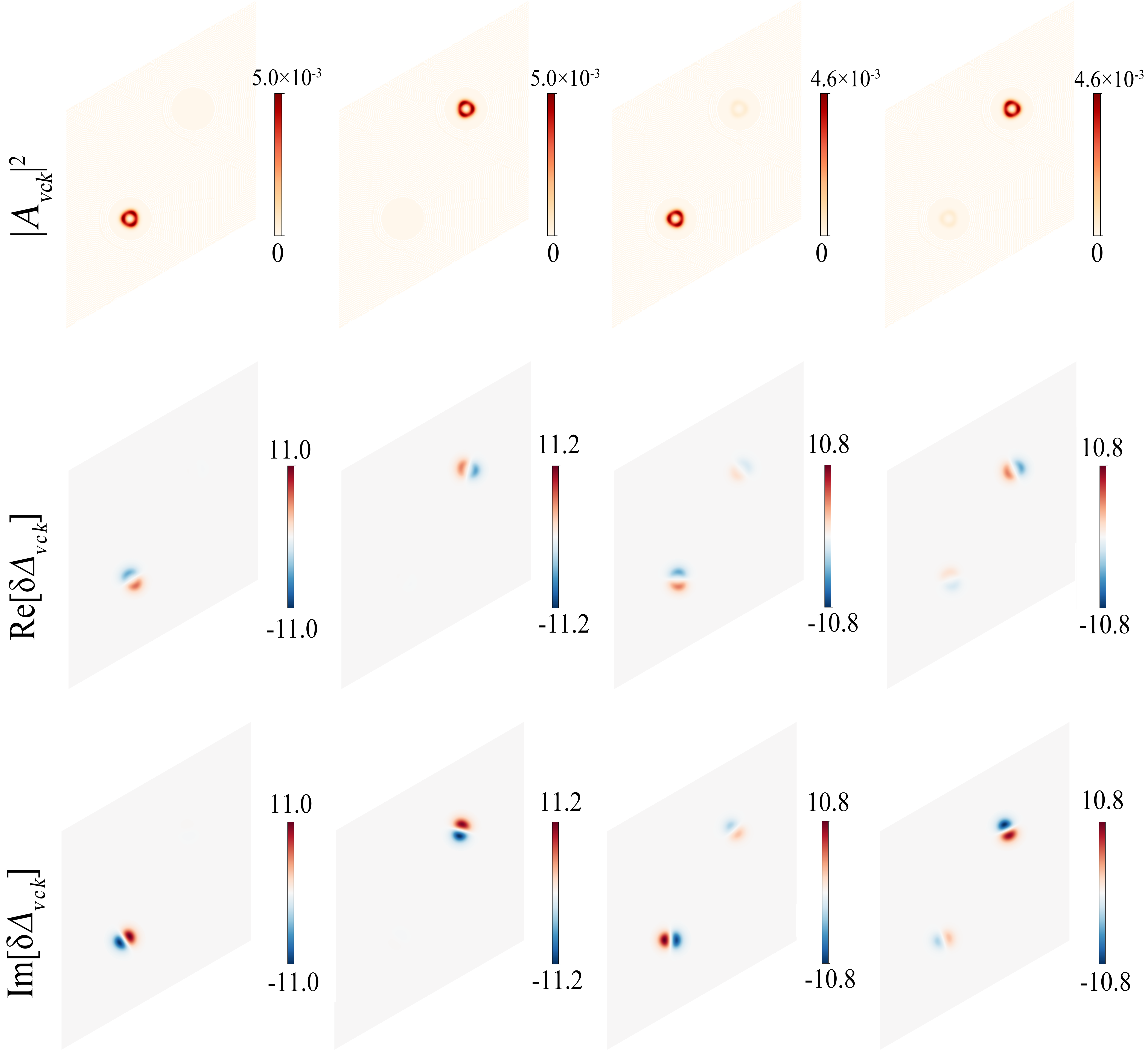}
\caption{Plot of $|A_{vck}|^2$ and the real and imaginary parts of $\delta \Delta_{vck}$ (meV) for all $2p$ modes of $1V1C$-configuration including both $K$ and $K'$ valleys.}
\label{KKp}
\end{figure*}
\begin{figure*}[h!]
\centering
\includegraphics[width=12.0cm,clip=true]{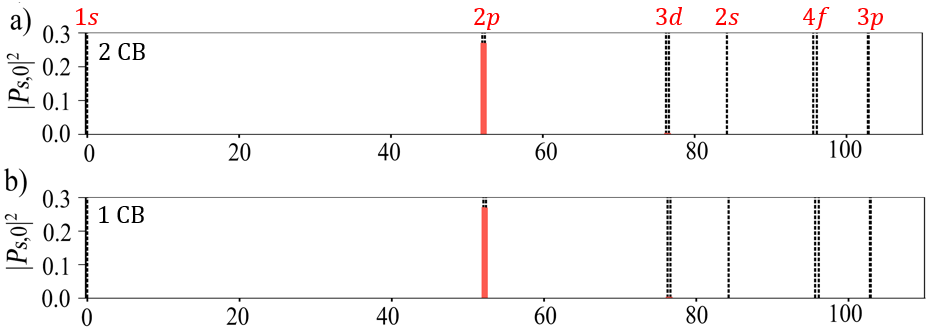}
\caption{Band convergence for the BSE calculated excitation spectrum with 2 (a) and 1 (b) conduction bands using the system without SOC effect. (b) is the same as Fig.\ref{vc}(b).}
\label{band}
\end{figure*}
\end{widetext}

\end{document}